\documentclass[12pt,a4paper]{article}
\pdfoutput=1
\usepackage{jcappub}
\usepackage{amsmath}
\usepackage{latexsym}
\usepackage{amsfonts}
\usepackage{graphicx}
\usepackage{mathrsfs}
\usepackage{epstopdf}
\usepackage{subfigure}

\title{On the constant-roll inflation}
\author{Zhu Yi}
\author[1]{and Yungui Gong\note{Corresponding author}}

\affiliation{School of Physics, Huazhong University of Science and Technology,
Wuhan, Hubei 430074, China}

\emailAdd{yizhu92@hust.edu.cn}
\emailAdd{yggong@mail.hust.edu.cn}

\abstract{
The primordial power spectra of scalar and tensor perturbations during slow-roll inflation are usually
calculated with the method of Bessel function approximation. For constant-roll or ultra slow-roll inflation,
the method of Bessel function approximation may be invalid. We compare the numerical results with the analytical results
derived from the Bessel function approximation, and we find that they differ significantly on super-horizon scales
if the constant slow-roll parameter $\eta_H$ is not small. More accurate method is needed for
calculating the primordial power spectrum for constant-roll inflation.}

\arxivnumber{1712.07478}

\begin{document}
\maketitle
\section{Introduction}
\label{sec-1}

Inflation gives solutions to the problems such as the flatness and horizon problems in standard cosmology \cite{Guth:1980zm,Linde:1981mu,Albrecht:1982wi,Starobinsky:1980te,Sato:1980yn}, and the seeds of the large scale structure of our Universe
are provided by the quantum fluctuations of the inflaton during inflation which leave imprints on the cosmic microwave background radiation
 \cite{Mukhanov:1981xt,Guth:1982ec,Hawking:1982cz,Bardeen:1983qw,Mukhanov:1985rz,Sasaki:1986hm}.
To ensure enough inflation, the potential of the scalar field is usually chosen to be flat
so that the scalar field slowly rolls down the potential,
i.e., the slow-roll condition is satisfied. Under the slow-roll condition,
the equation that the mode functions of the quantum fluctuations satisfy can be written as the Bessel equation with
the order $\nu$ and the order parameter $\nu$ is approximated to the first order of slow-roll parameters $\epsilon_H$ and $\eta_H$.
By assuming that the slow-roll parameters change slowly and taking $\nu$ as a constant, the mode functions are solved with
the Hankel functions through the matching of the solutions in different regions.
Finally the power spectra are evaluated at the horizon crossing $k=aH$
using the asymptotic form of the Hankel function in the limit $k/aH\rightarrow 0$ \cite{Stewart:1993bc}.
This method is referred as the Bessel function approximation \cite{Wang:1997cw}.
Higher order corrections and more accurate methods are also developed in \cite{Gong:2001he,Schwarz:2001vv,Habib:2002yi,Habib:2004kc}.
There are more work discussing the calculation of the power spectrum \cite{Lidsey:1995np,Easther:1995pc,Grivell:1996sr,Kinney:1997ne,Seto:1999jc,Adams:2001vc,Inoue:2001zt,
Tsamis:2003px,Kinney:2005vj,Zhu:2013upa,Hirano:2016gmv}.

If the potential of the scalar field is very flat, then the acceleration of the inflaton is locked by the friction term and
the slow-roll parameter $\eta_H\approx 3$, this is the ultra slow-roll inflation \cite{Tsamis:2003px,Kinney:2005vj}.
Recently, the ultra slow-roll inflation was generalized to constant-roll inflation with the slow-roll parameter $\eta_H$ being a constant \cite{Martin:2012pe,Motohashi:2014ppa}.
The constant-roll inflation has richer physics than the slow-roll inflation does.
For example, it can generate large local non-Gaussianity
and the curvature perturbation may grow on the super-horizon scales \cite{Martin:2012pe,Motohashi:2014ppa,Namjoo:2012aa}.
The constant-roll inflation was also discussed in \cite{Motohashi:2017vdc,Motohashi:2017aob,Nojiri:2017qvx,Oikonomou:2017bjx,
Odintsov:2017qpp,Dimopoulos:2017ged,Gao:2017owg,Ito:2017bnn,Karam:2017rpw,Cicciarella:2017nls,Anguelova:2017djf,Gao:2018tdb,Gao:2018cpp}.
Furthermore, the idea of ultra slow-roll inflation was used to generate the primordial black holes \cite{Germani:2017bcs,Gong:2017qlj},
and a short period of ultra slow-roll inflation with small $\epsilon_H$ can be
the mechanism for producing the primordial black holes in a similar way to
the chaotic new inflation model \cite{Saito:2008em}.
If $\eta_H\ll 1$, then the slow-roll conditions are satisfied and the standard slow-roll results apply.
If $\eta_H$ is not small, then the slow-roll approximation breaks down.
In the ultra slow-roll inflation, the slow-roll condition is violated.
As pointed out in \cite{Wang:1997cw}, when the slow-roll parameters are not small, the Bessel function approximation is unreliable.
Fortunately, since $\eta_H$ is a constant, we can separate it from the slow-roll parameter $\epsilon_H$
and use the Bessel function approximate to derive the power spectrum. It was shown
that the power spectrum for the constant-roll (including the ultra slow-roll) inflation has the same form as
that for the slow-roll inflation \cite{Kinney:2005vj,Martin:2012pe,Motohashi:2014ppa}.
One of the important assumptions made in the Bessel function approximation is that $\epsilon_H$ changes slowly.
For the slow-roll inflation, the time derivative of $\epsilon_H$ is
on the second order of the slow-roll parameters $\epsilon_H$ and $\eta_H$, but for
the constant-roll inflation, the time derivative of $\epsilon_H$ is
on the order of $\epsilon_H$, so the accuracy of the Bessel function approximation needs to be examined.
In this paper, we study this problem by comparing the power spectrum derived from Bessel function approximation with the numerical results.

This paper is organized as following. In section \ref{sec-2}, we briefly review the constant-roll inflation
and discuss the attractor behaviour.
In section \ref{sec-3}, we calculate the power spectrum using the Bessel function approximation method and compare the analytical results with the numerical results.
The issue of super-horizon evolution is also discussed in  section \ref{sec-3}.
The paper is concluded in section \ref{sec-5}.

\section{the constant-roll inflation}
\label{sec-2}
We start with the canonical scalar filed minimally coupled to gravity,
\begin{equation}
\label{action1}
S=\int d^4 x\sqrt{-g}\left[\frac{R}{2}-\frac12g^{\mu\nu}\nabla_\mu\phi\nabla_\nu\phi-V\left(\phi\right)\right],
\end{equation}
where we set $8\pi G=1$.
The Friedmann equation and the equation of motion for the scalar field are
\begin{gather}
\label{eq1}
3H^2=\frac{\dot{\phi}^2}{2}+V(\phi),\\
\label{phi:H}
\frac{d H}{d\phi}=-\frac{\dot{\phi}}{2},\\
\label{eq3}
\ddot{\phi}+3H\dot{\phi}+\frac{\partial V}{\partial \phi}=0,
\end{gather}
where a dot denotes the derivative with respect to cosmic time $t$. Introducing the slow-roll parameters
\begin{gather}
\label{sr1}
\epsilon_H=-\frac{\dot{H}}{H^2},\\
\label{sr2}
\eta_H=-\frac{\ddot{H}}{2H\dot{H}},
\end{gather}
and using eq. \eqref{phi:H}, we get
\begin{gather}
\label{sr3}
\epsilon_H=\frac{2}{H^2}\left(\frac{dH}{d\phi}\right)^2,\\
\label{sr4}
\eta_H=-\frac{\ddot{\phi}}{H\dot{\phi}}=\frac{2}{H}\frac{d^2H}{d\phi^2}.
\end{gather}
Note that the condition for inflation is $\epsilon_H\le 1$.
For the constant-roll inflation,
\begin{equation}
\label{constrl1}
\eta_H=-\frac{\ddot{\phi}}{H\dot{\phi}}=2\alpha.
\end{equation}
From equation \eqref{sr4}, we get
\begin{equation}
\label{crheq1}
\frac{1}{H}\frac{d^2H}{d\phi^2}=\alpha.
\end{equation}
If the constant $\alpha>0$,  then the solution of the Hubble parameter is
\begin{equation}\label{hubble}
  H\left(\phi\right)=c_1\exp\left[\sqrt{\alpha}\left(\phi-\phi_0\right)\right]+
  c_2\exp\left[-\sqrt{\alpha}\left(\phi-\phi_0\right)\right],
\end{equation}
where $c_1$, $c_2$ and $\phi_0$ are arbitrary integration constants.
For any values of $c_1$ and $c_2$, we can choose the value of $\phi_0$ so that
the solution \eqref{hubble} falls into one of the following three classes \cite{Motohashi:2014ppa}:
\begin{gather}
\label{exp1}
H(\phi)=M\exp(\pm\sqrt{\alpha}\phi),\quad c_1c_2=0,\\
\label{cosh1}
H(\phi)=M \cosh\left(\sqrt{\alpha} \phi \right),\quad c_1c_2>0,\\
\label{sinh0}
H(\phi)=M\sinh\left(\sqrt{\alpha}\phi\right),\quad c_1c_2<0,
\end{gather}
where the energy scale $M>0$.

On the other hand, if $\alpha<0$, the solution to equation \eqref{crheq1} is
\begin{equation}
\label{sin:hub}
H(\phi)=c_1\sin\left(\sqrt{-\alpha}\phi\right)+
  c_2\cos\left(\sqrt{-\alpha}\phi\right).
\end{equation}
By shifting the scalar field, the Hubble parameter \eqref{sin:hub} can be written as
\begin{equation}
\label{sin:hub2}
H(\phi)=M\sin\left[\sqrt{-\alpha} (\phi-\phi_1)\right],
\end{equation}
where $M>0$. The solution \eqref{sin:hub2} can also be expressed in the form of eq. \eqref{sinh0}, so
the general solution belongs to one of the three classes \eqref{exp1}, \eqref{cosh1} and \eqref{sinh0}.

The solution \eqref{exp1} leads to the power-law inflation, and $\epsilon_H=\eta_H=2\alpha$.
For the solution \eqref{cosh1}, we get
\begin{equation}
\label{cosh2}
\epsilon_H=2\alpha\tanh^2(\sqrt{\alpha}\phi),
\end{equation}
and
\begin{equation}
\label{cosh3}
\dot{\epsilon}_H=2H\epsilon_H(\epsilon_H-\eta_H)=-\frac{4\alpha H\epsilon_H}{\cosh^2(\sqrt{\alpha}\phi)}<0.
\end{equation}
For the models \eqref{exp1} and \eqref{cosh1}, the slow-roll parameter $\epsilon_H$ either is a constant or decreases with time,
so we need to introduce some mechanisms to end inflation.
The three classes of solutions \eqref{exp1}, \eqref{cosh1} and \eqref{sinh0} were discussed in \cite{Motohashi:2014ppa},
and it was claimed that the solution \eqref{sinh0} does not give inflationary phase. In this paper,
we focus on the solution \eqref{sinh0} only.

\begin{figure}[htbp]
  \centering{
  \includegraphics[width=0.5\textwidth]{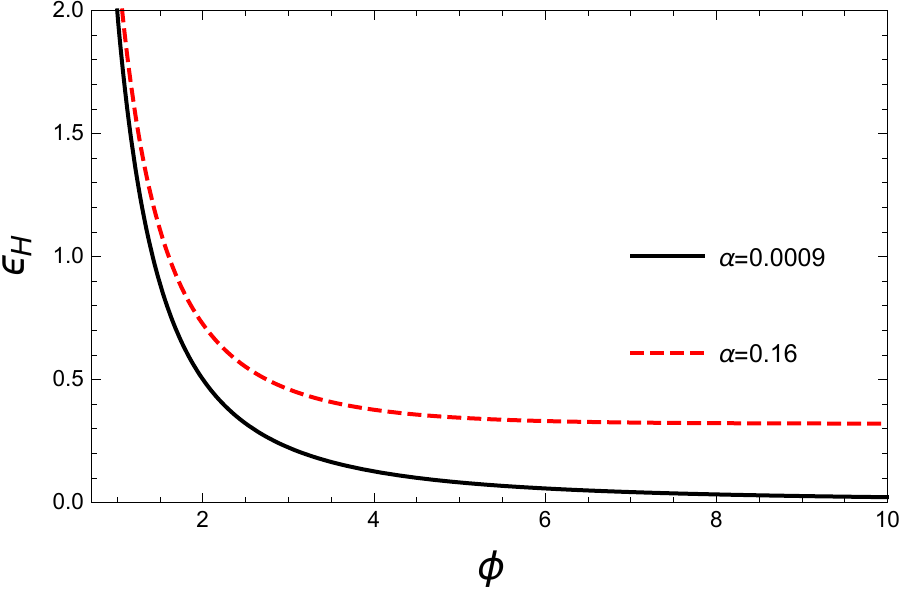}}
  \caption{The evolution of the slow-roll parameter $\epsilon_H$ for the model \eqref{sinh0}.}
  \label{epsilon_h}
\end{figure}

For the solution \eqref{sinh0}, we get
\begin{equation}
\label{sinh2}
\epsilon_H=2\alpha\left[\coth\left(\sqrt{\alpha}\phi\right)\right]^2,
\end{equation}
and
\begin{equation}
\label{sinh3}
\dot{\epsilon}_H=\frac{4\alpha H\epsilon_H}{\sinh^2\left(\sqrt{\alpha}\phi\right)}>0.
\end{equation}
Apparently, in this model inflation happens when $\epsilon_H<1$ and the slow-roll parameter $\epsilon_H$ increases with time as shown in figure \ref{epsilon_h}.
From eq. \eqref{cosh3}, we see that $\epsilon_H>\eta_H$ in this model since $\dot{\epsilon}_H>0$.
Note that for the model \eqref{sin:hub2} with $\alpha<0$, we also have $\dot{\epsilon}_H>0$.
From the Hamilton-Jacobi equation
\begin{equation}
\label{hj}
V(\phi)=3H^2(\phi)-2\left[\frac{d H\left(\phi\right)}{d\phi}\right]^2,
\end{equation}
and the solution \eqref{sinh0}, we obtain the potential
\begin{equation}
\label{p:sinh_h}
V(\phi)=V_0\sinh ^2\left(\sqrt{\alpha } \phi\right)-V_1 \cosh ^2\left(\sqrt{\alpha } \phi\right),
\end{equation}
where $V_0=3 M^2$ and $V_1=2 \alpha  M^2$.
Furthermore, we can derive the background evolution with the solution \eqref{sinh0}.
Combining eqs. \eqref{phi:H} and \eqref{sinh0}, we get
\begin{gather}
\label{phi_t}
\phi=\frac{1}{\sqrt{\alpha}}\ln \left(\tan\left[\alpha M\left(t_0-t\right)\right]\right),
\end{gather}
where $t_0$ is an integral constant and solution is valid in the range
\begin{equation}
\label{para:area}
\pi/4<\alpha M\left(t_0-t\right)<\pi/2.
\end{equation}
Combining eqs. \eqref{sinh0} and \eqref{phi_t}, we get the time evolution of the scale factor
\begin{equation}\label{scale}
a=a_0\sin^{1/(2\alpha)}[2\alpha M(t_0-t)],
\end{equation}
where $a_0$ is an integration constant.

From eq. \eqref{sinh2} and the condition for the end of inflation $\epsilon_H(\phi_e)=1$, we get
the value of the inflaton $\phi_e$ at the end of inflation,
\begin{equation}\label{sinhphie}
    \phi_e=\frac{1}{2\sqrt{\alpha}}\ln\left(\frac{1+\sqrt{2\alpha}}{1-\sqrt{2\alpha}}\right),
\end{equation}
and the value of the inflaton $\phi_*$ at the horizon exit,
\begin{equation}\label{sinhphii}
   \phi_*=\frac{1}{\sqrt{\alpha}}\ln\left(\frac{y+\sqrt{y^2-4}}{2}\right),
\end{equation}
where $\alpha<1/2$, $y=\exp\left(2\alpha N+\sqrt{\alpha}\phi_e\right)+\exp\left(2\alpha N-\sqrt{\alpha}\phi_e\right)$,
and $N$ is the remaining number of $e$-folds at the horizon exit before the end of inflation. Since $\alpha<1/2$, so
to get successful inflation, we require $\eta_H=2\alpha<1$ for the model \eqref{sinh0}.
Substituting the result \eqref{sinhphii} into eq. \eqref{sinh2}, we get the value of the
slow-roll parameter at the horizon exit,
\begin{equation}
\label{epsion1}
\epsilon_H(\phi_*)=2\alpha\left(1+\frac{1}{1+\sqrt{1-N_*^{-1}}-N_*^{-1}}\times\frac{1}{N_*}\right)^2,
\end{equation}
where $N_*=\exp(4\alpha N)/(1-2\alpha)$.

\begin{figure}
$\begin{array}{cc}
\includegraphics[width=0.4\textwidth]{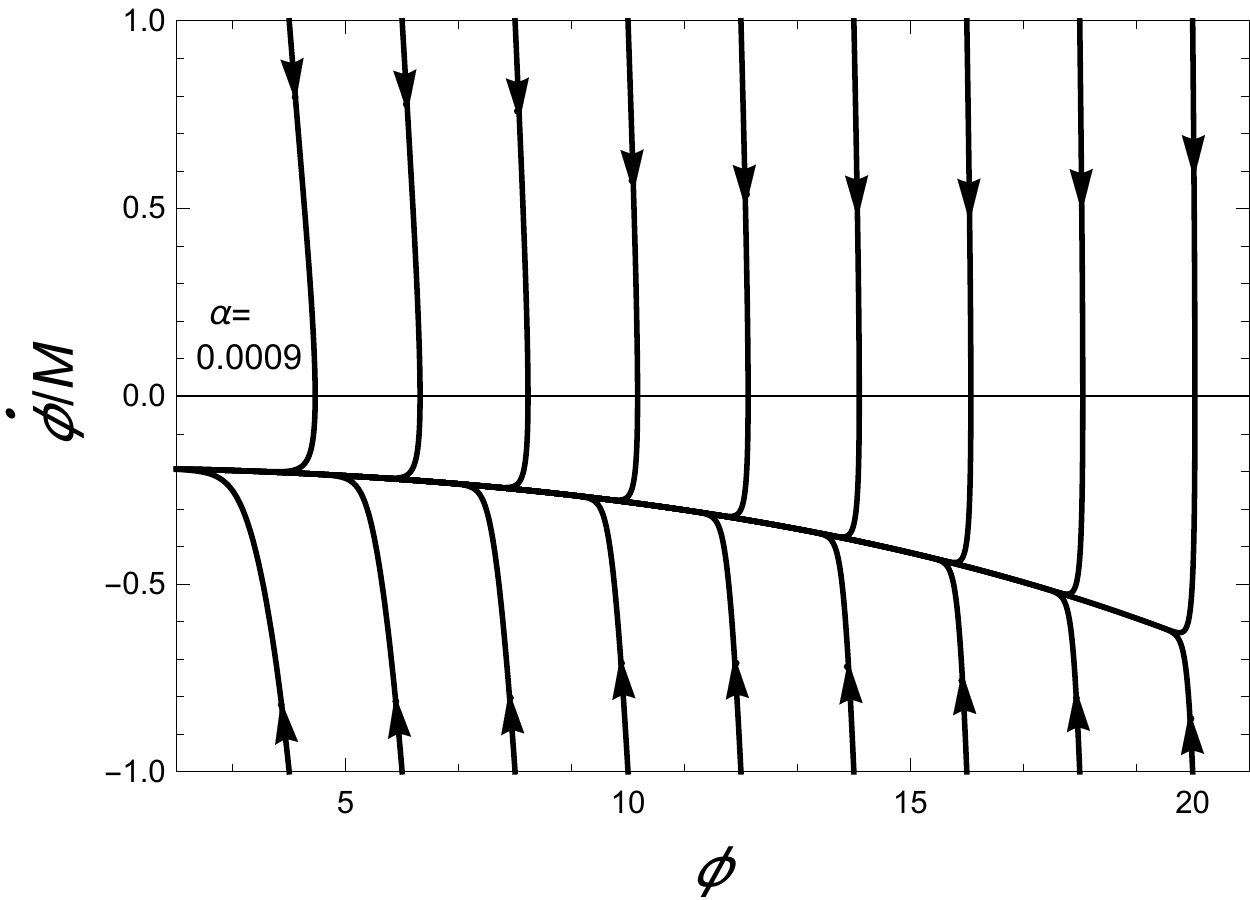}&\qquad
\includegraphics[width=0.4\textwidth]{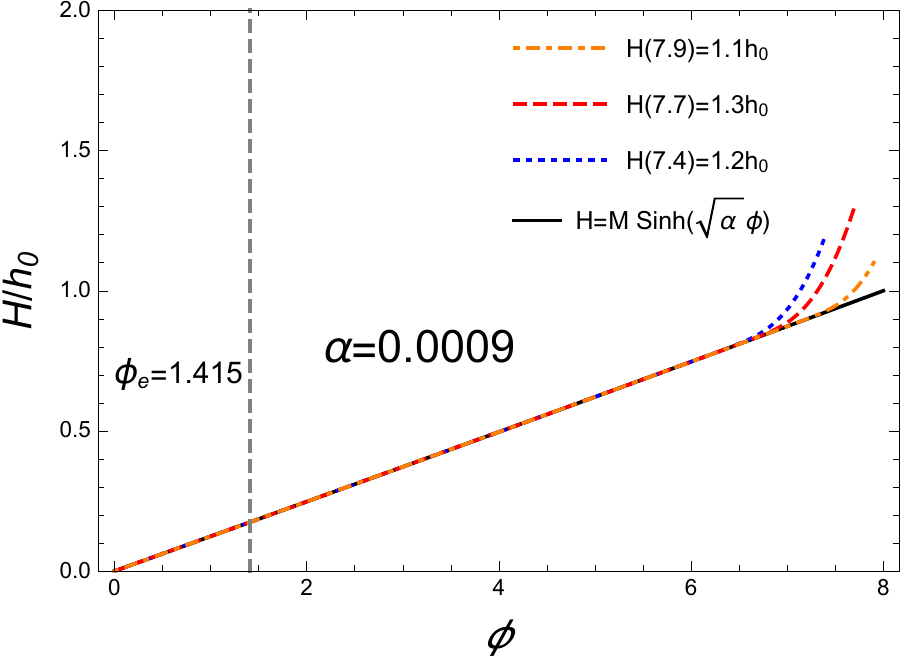}\\
\includegraphics[width=0.4\textwidth]{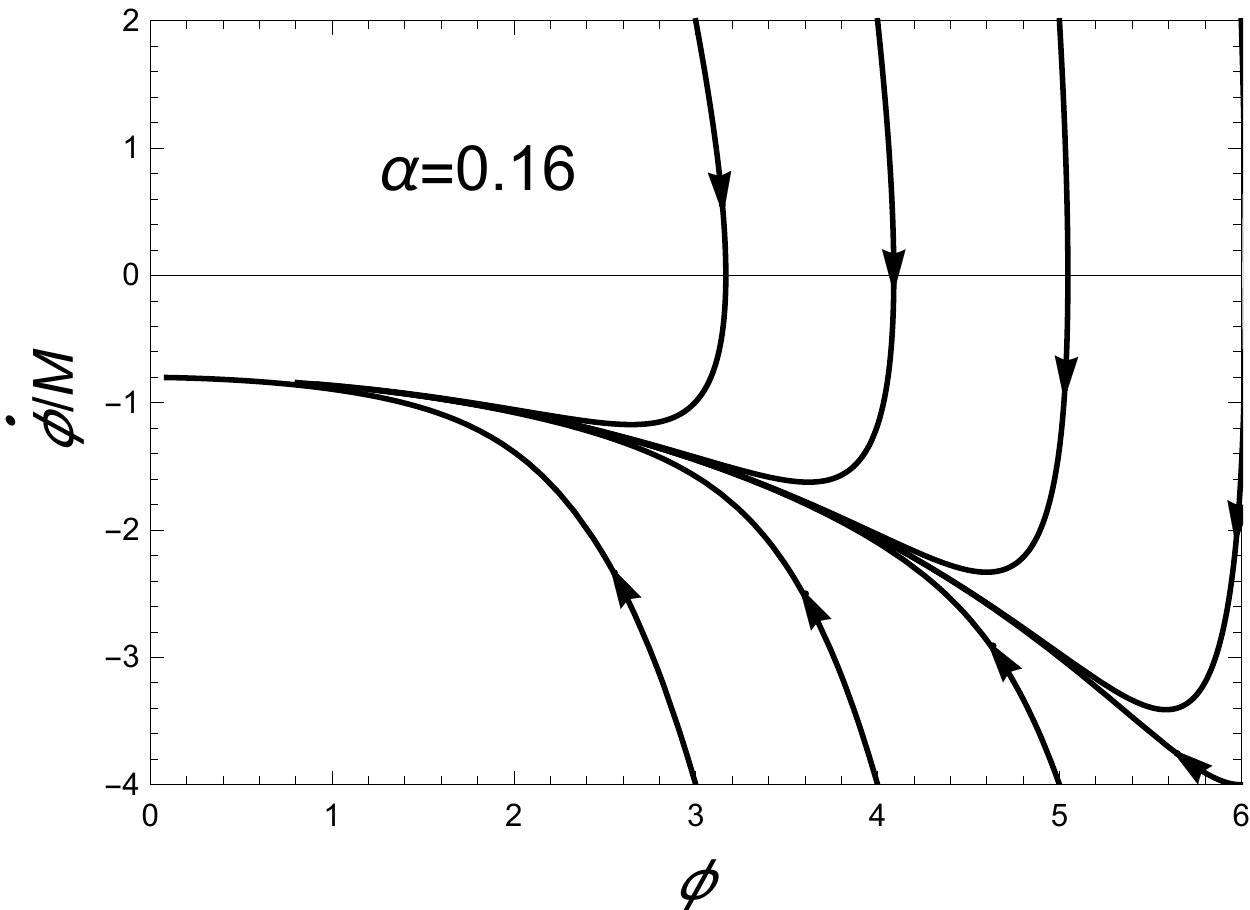}&\qquad
\includegraphics[width=0.4\textwidth]{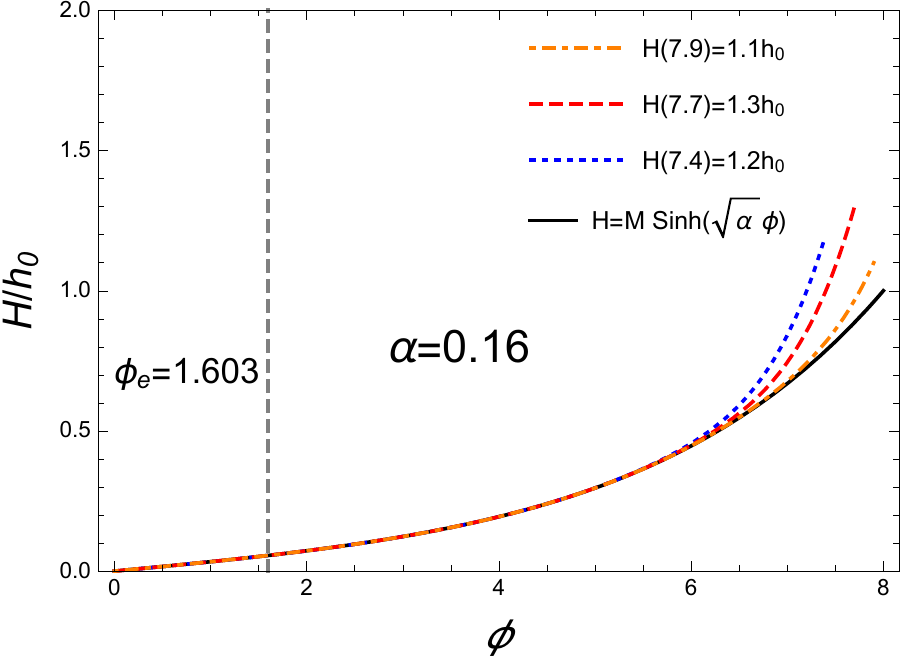}\\
\end{array}$
\caption{The phase diagram for the model \eqref{sinh0} with $\alpha=0.0009$ (top) and $\alpha=0.16$ (bottom).
In the right panels, the Hubble parameters are normalized by $h_0=M\sinh(8\sqrt{\alpha})$.}
\label{fig2}
\end{figure}

To close this section, we discuss whether the exact inflationary trajectories \eqref{phi_t} and \eqref{scale} are attractors.
We solve eqs. \eqref{eq1} and \eqref{eq3} together with the Hamilton-Jacobi equation \eqref{hj}
for the model \eqref{sinh0} numerically with different initial conditions and the results are shown in figure \ref{fig2}.
The results show that the inflationary trajectories are attractors.

\section{perturbation in the constant-roll inflation}
\label{sec-3}
\subsection{The scalar perturbation}
\label{subsec-scalar}
In the flat gauge $\delta\phi=0$, the gauge invariant scalar perturbation becomes the curvature perturbation
which is related to the scalar metric perturbation by $\delta g_{ij}=a^2(1+2\zeta)\delta_{ij}$.
The mode function $v_k=z\zeta_k$ for the curvature perturbation $\zeta$ satisfies the
Mukhanov-Sasaki equation \cite{Mukhanov:1985rz,Sasaki:1986hm}
\begin{equation}
\label{mseq}
v_k''+\left(k^2-\frac{z''}{z}\right)v_k=0,
\end{equation}
where $z=a\dot{\phi}/H$ and a prime denotes the derivative with respect to the conformal time $\tau$.
In terms of the slow-roll parameters, we get
\begin{equation}
\label{zz1}
  \frac{z''}{z}=2a^2H^2\left(1+\epsilon_H-\frac32\eta_H+\epsilon_H^2+\frac12 \eta_H^2-2\epsilon_H\eta_H+\frac12 \xi_H\right),
\end{equation}
where
\begin{equation}
  \xi_H=\frac{\dddot{\phi}}{H^2\dot{\phi}}-\eta_H^2.
\end{equation}
Since
\begin{equation}
\label{deta_H}
\dot{\eta}_H=H\left(\epsilon_H\eta_H-\xi_H\right),
\end{equation}
so for the constant-roll inflation with $\dot{\eta}_H=0$,
we obtain $\xi_H=\epsilon_H\eta_H$, and
eq. \eqref{zz1} becomes
\begin{equation}
\label{zz2}
\frac{z''}{z}=2a^2H^2\left(1+\epsilon_H-\frac32\eta_H+\epsilon_H^2+\frac12 \eta_H^2-\frac32\epsilon_H\eta_H\right).
\end{equation}
To find the relation between $aH$ and $\tau$, we use the relation
\begin{equation}
\label{tau:epsilon}
\frac{d}{d\tau}\left(\frac{1}{aH}\right)=-1+\epsilon_H.
\end{equation}
To the first order approximation of $\epsilon_H$, we get
\begin{equation}
\label{tau}
\frac{1}{aH}\approx \left(-1+\frac{\epsilon_H}{1+2\eta_H}\right)\tau.
\end{equation}
If $\epsilon_H$ is almost a constant or not small, then this approximation is not correct.
Plugging the above result \eqref{tau} into eq. \eqref{zz2}, we get
\begin{equation}
\label{zz3}
\begin{split}
\frac{z''}{z}&=\frac{1}{\tau^2}\left[\left(2-3\eta_H+\eta_H^2\right)
  +\frac{\left(6-5\eta_H-4\eta_H^2\right)\epsilon_H}{1+2\eta_H}\right]\\
&=\frac{1}{\tau^2}\left(\nu^2-\frac14\right),
\end{split}
\end{equation}
where to the first order approximation of $\epsilon_H$,
\begin{equation}
\label{nu:value}
\nu=\left|\frac32-\eta_H\right|
  +\frac{\left(6-5\eta_H-4\eta_H^2\right)\epsilon_H}{\left|3-2\eta_H\right|\left(1+2\eta_H\right)}.
\end{equation}
In \cite{Kinney:2005vj,Martin:2012pe}, they got the zeroth order result $\nu=|3-2\eta_H|/2$ only,
here we extend the result to the first order of $\epsilon_H$ with the help of the more accurate relation \eqref{tau}.
Assuming that $\nu$ is a constant, we get the solution to eq. \eqref{mseq}
\begin{equation}
\label{Hankel1}
v_k=\sqrt{-\tau}\left[b_1H_{\nu}^{~(1)}\left(-k\tau\right)+b_2H_{\nu}^{~(2)}\left(-k\tau\right)\right],
\end{equation}
where $b_1$ and $b_2$ are arbitrary integration constants. By choosing the Bunch-Davies vacuum and matching the
asymptotic plane wave solution
\begin{equation}
\label{boundarycondtion1}
v_k\left(\tau\right)=\frac{1}{\sqrt{2k}}\exp\left(-i k\tau\right),
\end{equation}
for $-k\tau\gg 1$, we get the mode function
\begin{equation}
\label{Hankel2}
v_k=\frac{\sqrt{\pi}}{2}\exp\left[i\left(\nu+\frac12\right)\frac{\pi}{2}\right]\sqrt{-\tau}H_{\nu}^{(1)}\left(-k\tau\right).
\end{equation}
Substituting eq. \eqref{tau} into eq. \eqref{Hankel2}, we get the power spectrum for the scalar perturbation
\begin{equation}
\label{perturbation1}
\mathcal{P_{R}}=\frac{k^3}{2\pi^2}\left|\zeta_k\right|^2=\frac{H^2}{16\pi\epsilon_H }\frac{1+2\eta_H}{1+2\eta_H-\epsilon_H}
\left[H_{\nu}^{(1)}\left(\frac{1+2\eta_H}{1+2\eta_H-\epsilon_H}\frac{k}{aH}\right)\right]^2\left(\frac{k}{aH}\right)^3.
\end{equation}
On super-horizon scales, $k\ll aH$, using the asymptotic behaviour of the Hankel function, the power spectrum for the scalar perturbation becomes
\begin{equation}
\label{perturbation12}
\mathcal{P_{R}}=2^{2\nu-3}\left[\frac{\Gamma\left(\nu\right)}{\Gamma\left(3/2\right)}\right]^2
  \frac{1}{2\epsilon_H}\left(\frac{H}{2\pi}\right)^2\left(1-\frac{\epsilon_H}{1+2\eta_H}\right)^{2\nu-1}\left(\frac{k}{aH}\right)^{3-2\nu}.
\end{equation}
The expression is almost the same as that for the slow-roll inflation except that the value of $\nu$ is different.
Similar to the slow-roll inflation, we get the scalar spectral tilt
\begin{equation}
\label{nseq1}
n_s-1=\frac{d\ln\mathcal{P_{R}}}{d\ln k}=3-2\nu.
\end{equation}
Note that the above results are valid for any constant-roll inflation.
To the zeroth order of $\epsilon_H$, we recover the result obtained in \cite{Kinney:2005vj,Martin:2012pe}.
Combining eqs. \eqref{constrl1}, \eqref{epsion1}, \eqref{nu:value} and \eqref{nseq1}, we get
\begin{equation}
\label{nseq3}
n_s-1=4\alpha\left[1-\frac{(6-10\alpha-16\alpha^2)}{(3-4\alpha)(1+4\alpha)}\left(1+\frac{1}{1+\sqrt{1-N_*^{-1}}-N_*^{-1}}\times\frac{1}{N_*}\right)^2\right].
\end{equation}

\subsection{Super-horizon evolution}
For the constant-roll inflation, the curvature perturbation $\zeta$ may evolve on the super-horizon scales \cite{Namjoo:2012aa,Martin:2012pe,Motohashi:2014ppa}.
In this section, we discuss the evolution of the curvature perturbation  $\zeta$ on the super-horizon scales for the model \eqref{sinh0}.
On super-horizon scales, the $k^2$ term in eq. \eqref{mseq} is negligible and the general solution is
\begin{equation}
\label{zeta1}
\zeta_k(\tau)=A_k+B_k \int^{\tau}\frac{1}{z^2(\bar\tau)}d\bar\tau,
\end{equation}
where the $k$-dependent constants $A_k$ and $B_k$ are determined by the initial conditions \eqref{boundarycondtion1}.
The first term $A_k$ gives the constant mode and the $B_k$ term gives the evolution of the curvature perturbation.
Changing the integral variable, the $B_k$ term becomes
\begin{equation}
\label{zeta0}
\zeta_0=\int^{\tau}\frac{d\bar\tau}{z^2(\bar\tau)}
=\int^{t}\frac{H^2}{a^3\dot{\phi}^2}d t.
\end{equation}
Substituting eqs. \eqref{phi_t} and \eqref{scale} into eq. \eqref{zeta0}, we obtain
\begin{equation}
\label{perturbation2}
\begin{split}
\zeta_0= & -\frac{1}{8\alpha^2 a_0^3 M}\int\frac{\cos^2\left[2\alpha M \left(t_0-t\right)\right]}{\sin^{3/2\alpha }\left[2\alpha M \left(t_0-t\right)\right]} d \left[2\alpha M \left(t_0-t\right)\right] \\
=& C x^{3} \, _2F_1\left(\frac32,\frac12+\frac{3}{4\alpha},\frac52,x^2\right),
\end{split}
\end{equation}
where $C=-(24\alpha^2 a_0^3 M)^{-1}$, $x=-\cos\left[2\alpha M \left(t_0-t\right)\right]$ and
$_2F_1(a,b;c;z)$ is the Hypergeometric function. On super-horizon scales, $2\alpha M \left(t_0-t\right)\rightarrow \pi/2$ and $x\ll 1$,
so
\begin{equation}
\label{perturbation32}
\zeta_k\approx A_k+B_k C x^{3}.
\end{equation}
The second term is a decaying mode and the constant mode dominates.
Therefore, the curvature perturbation $\zeta_k$ remains constant outside the horizon for the model \eqref{sinh0}
and we can calculate the value of the observable at the horizon exit $k=aH$.

\subsection{The tensor perturbation}
\label{subsec-ten}
\begin{figure}[htbp]
  \centering{
  \includegraphics[width=0.5\textwidth]{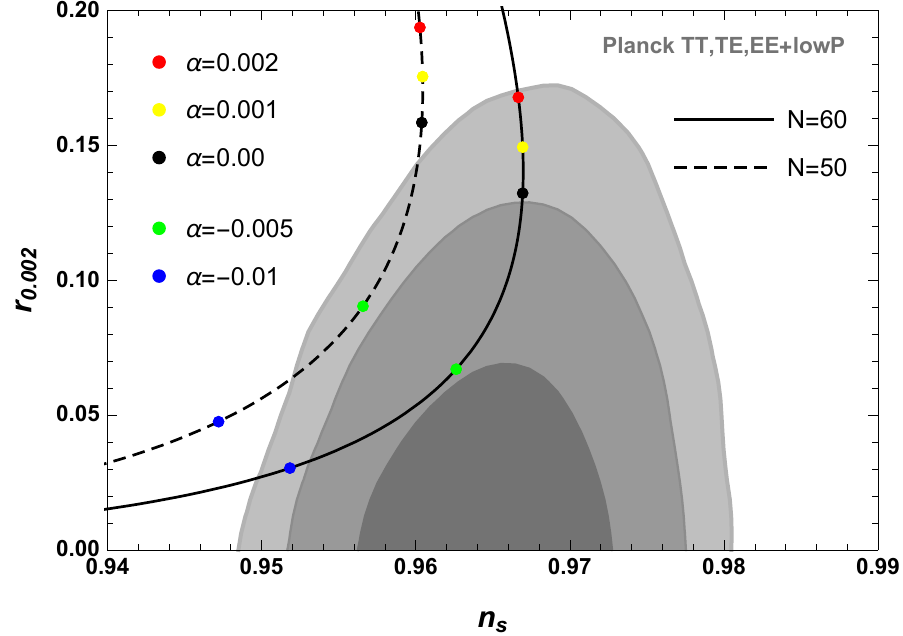}}
  \caption{The $n_s-r$ constraints on the model \eqref{sinh0}. The shaded regions are the marginalized 68\%, 95\% and 99.8\% C.L. contours from Planck 2015 data.}
  \label{obsnsr}
\end{figure}

For the tensor perturbation $\delta g_{ij}=a^2 h_{ij}$, the mode function $u_k^s(\tau)=a h^s_k(\tau)/\sqrt{2}$
satisfies the equation
\begin{equation}
\label{ten:per}
  \frac{d^2u_k^s}{d\tau^2}+\left(k^2-\frac{a''}{a}\right)u_k^s=0,
\end{equation}
where ``$s$" stands for the ``$+$" or ``$\times$" polarizations. Using the approximation \eqref{tau}, we get
\begin{equation}
\label{a2a}
\frac{a''}{a}=\frac{1}{\tau^2}\left[2+\frac{3-2\eta_H}{1+2\eta_H}\epsilon_H\right]=\frac{\mu^2-1/4}{\tau^2},
\end{equation}
where
\begin{equation}
\label{mu2}
\mu=\frac32 +\frac{3-2\eta_H}{3\left(1+2\eta_H\right)}\epsilon_H.
\end{equation}
Assuming that $\mu$ is a constant and following the same procedure for the scalar perturbation,
we get the power spectrum for the tensor perturbation
\begin{equation}
\label{ten:solution}
\mathcal{P_{T}}=\frac{H^2}{\pi}\frac{1+2\eta_H}{1+2\eta_H-\epsilon_H}
\left[H_{\mu}^{(1)}\left(\frac{1+2\eta_H}{1+2\eta_H-\epsilon_H}\frac{k}{aH}\right)\right]^2\left(\frac{k}{aH}\right)^3.
\end{equation}
On super-horizon scales, using the asymptotic behaviour of the Hankel function, the power spectrum for the tensor perturbation becomes
\begin{equation}
\label{ten:solution02}
\mathcal{P_{T}}=2^{2\mu}\left[\frac{\Gamma\left(\mu\right)}{\Gamma\left(3/2\right)}\right]^2
\left(\frac{H}{2\pi}\right)^2\left(1-\frac{\epsilon_H}{1+2\eta_H}\right)^{2\mu-1}\left(\frac{k}{aH}\right)^{3-2\mu}.
\end{equation}
The tensor spectral tilt is
\begin{equation}
\label{nt}
n_t=3-2\mu=-\frac{6-8\alpha}{3\left(1+4\alpha\right)}\epsilon_H.
\end{equation}
Combing eqs. \eqref{perturbation12} and \eqref{ten:solution02}, we obtain the tensor to scalar ratio
\begin{equation}
\label{ratio1}
r=\frac{\mathcal{P_{T}}}{\mathcal{P_{R}}}\approx 16 Q \epsilon_H=-8Q\left(1+\frac{16\alpha}{3-4\alpha}\right) n_t,
\end{equation}
where $Q=2^{3-|3-4\alpha|}[\Gamma(3/2)/\Gamma(|3/2-2\alpha|)]^2$.
Using eqs. \eqref{nseq3} and \eqref{ratio1}, we calculate the scalar spectral tilt and the tensor to scalar ratio for different values of $\alpha$ with $N=50$ and $N=60$.
The results along with the Planck 2015 constraints \cite{Ade:2015lrj} are shown in figure \ref{obsnsr}.
From figure \ref{obsnsr}, we see that the model gives too large $r$ and it is inconsistent with the observations at the 68\% confidence level.
Note that when $\alpha$ is small as those values chosen in figure \ref{obsnsr} to be consistent
with the observations at the 95\% confidence level,
the constant-roll inflation is also slow-roll inflation,
so the formulae \eqref{nseq3} and \eqref{ratio1} are valid.

\subsection{Numerical solution}
\label{sec-4}
\begin{figure}[htbp]
\centering
\includegraphics[width=0.45\textwidth]{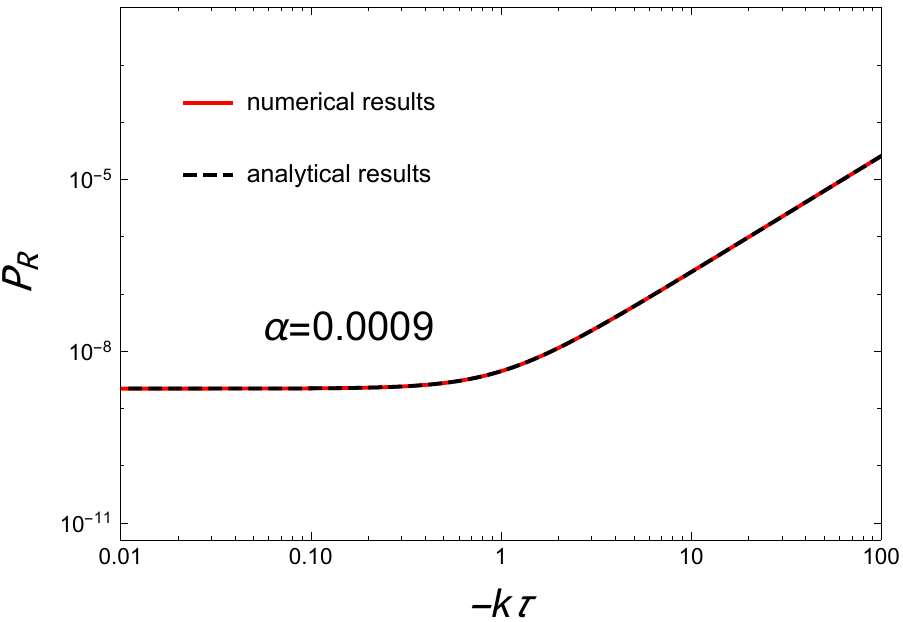}
\includegraphics[width=0.45\textwidth]{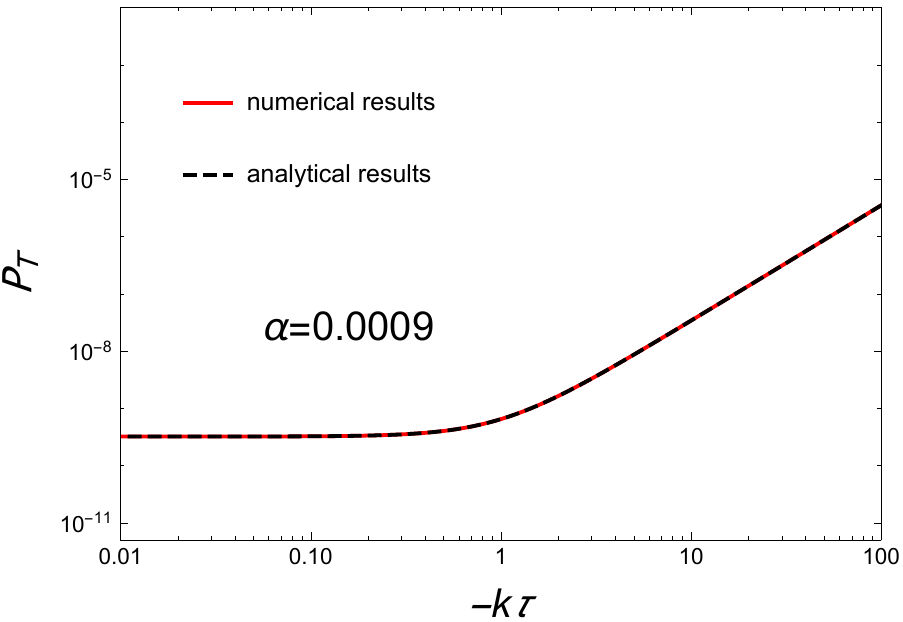}
\caption{The scalar and tensor power spectra for the constant-roll inflation with $\eta_H=0.0018$. The left panel shows the scalar power spectrum and the right panel shows
  the tensor power spectrum.  The solid red curves are for the numerical results and the black dashed curves are plotted with the analytical results \eqref{perturbation1} and \eqref{ten:solution}.}
\label{alpha01}
\end{figure}

In the derivation of the power spectra for the scalar and tensor perturbations, we made several approximations. The relation \eqref{tau}
between $aH$ and $\tau$ is an approximation to the first order of the slow-roll parameter $\epsilon_H$. To solve the mode functions, we assume
that $\mu$ and $\nu$ are constants. For the slow-roll inflation,
the time derivatives of $\mu$ and $\nu$ are at least on the second order of the slow-roll parameters $\epsilon_H$ and $\eta_H$
and the slow-roll parameters are small, so the error due to these approximations is small.
However, for the constant-roll inflation, $\dot\epsilon_H>0$, so $\epsilon_H>\eta_H$ and
the slow-roll parameter $\epsilon_H$ may not be too small as shown in figure \ref{epsilon_h}, and the time derivatives of $\mu$ and $\nu$ is in the first order of $\epsilon_H$,
so the error due to these approximations may not be small. To check the analytical results \eqref{perturbation1} and \eqref{ten:solution}
derived in the previous sections, we solve eqs. \eqref{mseq} and \eqref{ten:per} for the mode functions $v_k$ and $u_k^s$ numerically.
In figure \ref{alpha01}, we show the results for the case $\eta_H=0.0018$ with $k=0.002\text{Mpc}^{-1}$. For this case, the slow-roll condition is satisfied, the constant-roll
inflation is also a slow-roll inflation, so the theoretical results \eqref{perturbation1} and \eqref{ten:solution}
are almost the same as the numerical results.

\begin{figure}[htbp]
  \centering
  \includegraphics[width=0.45\textwidth]{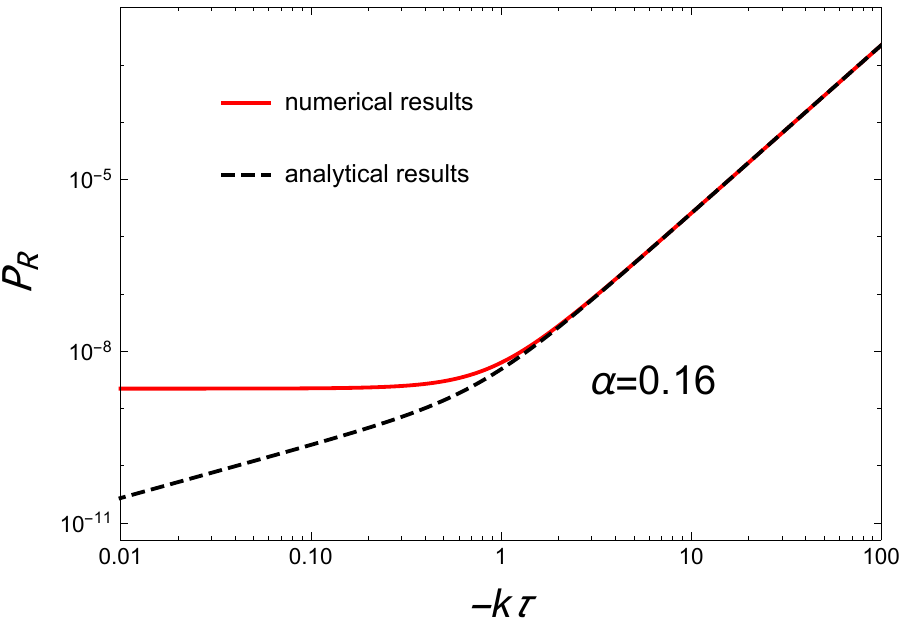}
  \includegraphics[width=0.45\textwidth]{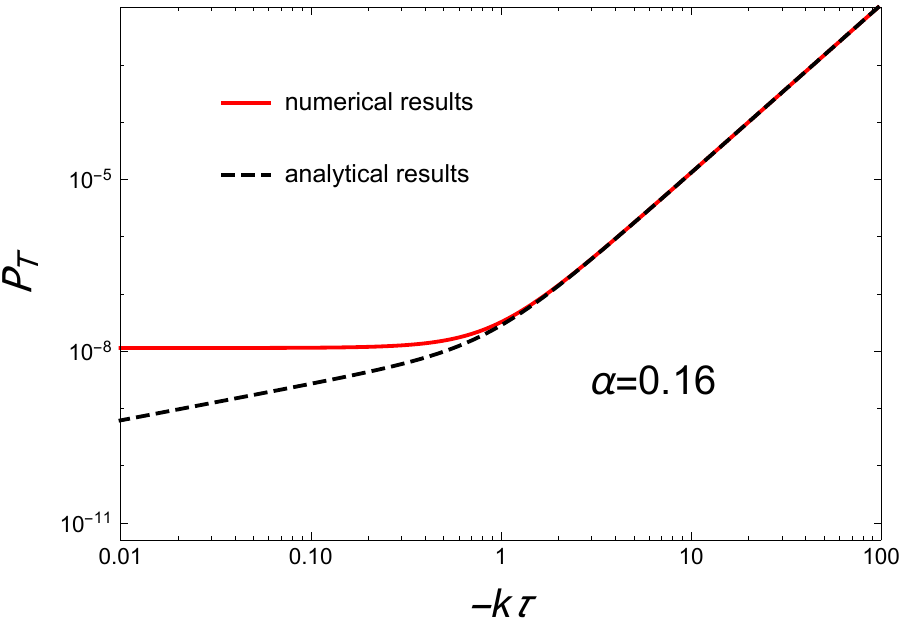}
\caption{The scalar and tensor power spectra for the constant-roll inflation with $\eta_H=0.32$. The left panel shows the scalar power spectrum and the right panel shows
the tensor power spectrum.  The solid red curves are for the numerical results and the black dashed curves are plotted with the analytical results \eqref{perturbation1} and \eqref{ten:solution}.}
\label{alpha04}
\end{figure}

\begin{figure}[htbp]
  \centering{
  \includegraphics[width=0.45\textwidth]{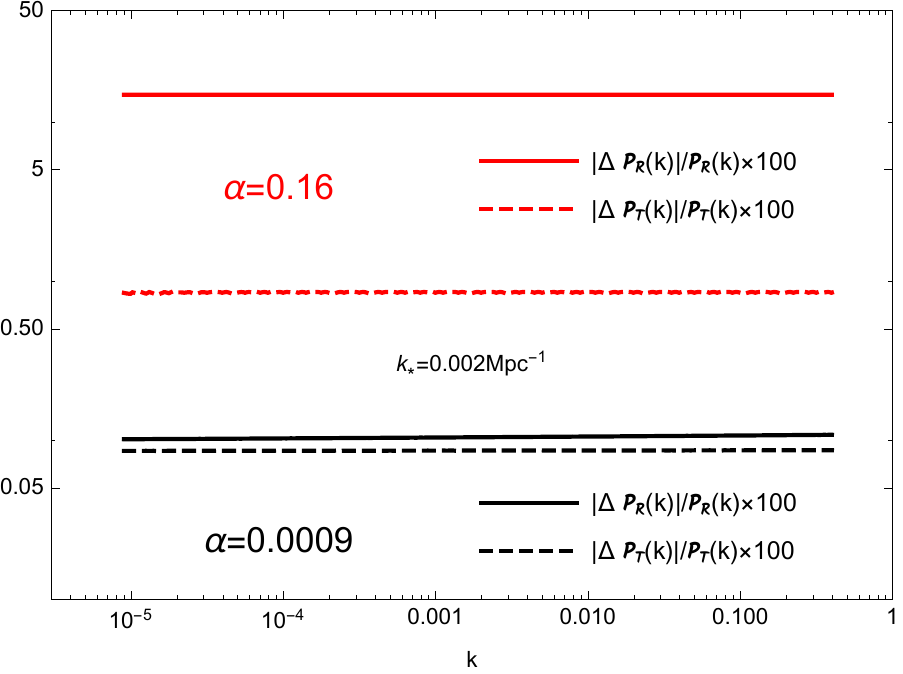}}
\caption{The percentage differences between the power spectra obtained from the numerical solutions and the analytical results \eqref{perturbation12}
and \eqref{ten:solution02}. The solid lines are for the scalar power spectrum and the dashed lines are for the tensor power spectrum.
The black lines are for the case with $\eta_H=0.0018$ and the red lines are for the case with $\eta_H=0.32$.}
\label{pwrdiff1}
\end{figure}

In figure \ref{alpha04}, we show the results for the constant-roll inflation with $\eta_H=0.32$.
For this case, the slow-roll parameters are not small,
the theoretical results \eqref{perturbation1} and \eqref{ten:solution}
differ from the numerical results significantly on super-horizon scales.
On sub-horizon scales $-k\tau\gg 1$, the asymptotic plane wave solution \eqref{boundarycondtion1} applies
and the solution is independent of the model,
so the theoretical and the numerical results are the same. On super-horizon scales,
$k$ is negligible, the solution depends on the inflationary model and the results are different for different models.
Figure \ref{alpha04} tells us that the analytical results \eqref{perturbation1} and \eqref{ten:solution} are not good approximate solutions
and we need to solve the equations for the mode functions numerically if $\eta_H$ is not small.
Furthermore, the numerical results confirm
that the perturbations remain constant on the super-horizon scales.
We also compare the difference of the power spectra between the numerical solutions and the analytical results \eqref{perturbation12}
and \eqref{ten:solution02}, the percentage difference is shown in figure \ref{pwrdiff1}.
As expected, the error in the scalar power spectrum is big when $\epsilon_H$ is not small.
However, the analytical results \eqref{perturbation12}
and \eqref{ten:solution02} capture the main behavior of $H^2/\epsilon_H$ even though they underestimate the amplitudes.
For the constant-roll inflation with $\eta_H=0.0018$, the analytical results give $n_s=0.9670$, $n_t=-0.0183$ and $r_{0.002}=0.1475$,
and the numerical results are $n_s=0.9666$, $n_t=-0.0187$ and $r_{0.002}=0.1455$.
For the constant-roll inflation with $\eta_H=0.32$, the analytical results give $n_s=0.98$, $n_t=-0.31$ and $r_{0.002}=7.34$,
and the numerical results are $n_s=0.059$, $n_t=-0.94$ and $r_{0.002}=5.12$.
Since further approximations were made in calculating the spectral tilts, the derived analytical results \eqref{nseq3}, \eqref{nt} and \eqref{ratio1} deviate from the exact numerical results when $\epsilon_H$ is not small.

\section{Discussions and conclusions}
\label{sec-5}

For the constant-roll inflation, we derived the power spectra for the scalar and tensor perturbations up
to the first order of $\epsilon_H$ by using the Bessel function approximation.
The scalar and tensor spectral tilts, and the tensor to scalar ratio
were also obtained up to the first order of $\epsilon_H$.
The reason that we can give the first order correction is because we derived the first order relation \eqref{tau}
for the constant-roll inflation. These extend the previous zeroth order result.
When $\eta_H$ is small, we recover the standard results for the slow-roll inflation.
The analytical inflationary solution for the constant-roll
model \eqref{sinh0} was also derived,
and we find that the inflationary trajectory is an attractor,
but the model is excluded by the observations at the 68\% C.L.
For the model \eqref{sinh0}, $\dot\epsilon_H>0$, so
as the inflaton rolls down the potential, $\epsilon_H>\eta_H$ and inflation ends when $\epsilon_H$ reaches 1. If $\eta_H$ is not small, then $\epsilon_H$ is not small and the Bessel function approximation may not be reliable. We solve the equations for the mode functions numerically and we find that the numerical results are different from the analytical ones derived from the Bessel function approximation outside the horizon.
Therefore, more accurate method for calculating the primordial power spectra is needed.
If $\dot{\epsilon}_H<0$, then $\epsilon_H$ can keep to be small and the Bessel function approximation is valid
if $\epsilon_H$ changes slowly and the curvature perturbation remains to be constant outside the horizon,
but inflation cannot end, and some mechanisms are needed to end inflation.
The small $\epsilon_H$ gives large curvature perturbation on small scales to form the primordial black holes,
so a short period of ultra slow-roll inflation with small $\epsilon_H$ can be the mechanism for producing the primordial black holes.
The results derived in this paper may be useful for the discussion of the production of primordial black holes.
In conclusion, the Bessel function approximation cannot be trusted if $\epsilon_H$ is not small or it changes quickly,
more accurate method for calculating the primordial power spectra and the spectral tilts needs to be developed.

\begin{acknowledgments}
This research was supported in part by the Natural Science
Foundation of China under Grant No. 11475065 and the Major Program of the National Natural Science Foundation of China under Grant No. 11690021.
\end{acknowledgments}

%\bibliographystyle{JHEP}
%\bibliography{../../book/cosmologyref}

\providecommand{\href}[2]{#2}\begingroup\raggedright\endgroup

\end{document}